\documentclass[9pt]{article}
\usepackage{spconf,amsmath,graphicx}
\usepackage{algorithm}
\usepackage{algpseudocode}
\usepackage{array}
\usepackage{bm}
\usepackage{soul}
\usepackage{subcaption}
\usepackage[sorting=none,url=true, giveninits=true]{biblatex}
\usepackage{amssymb}
\usepackage{tikz}
\title{Deep Reinforcement Learning for  IRS phase shift design in spatiotemporally correlated environments }
\name{Spilios Evmorfos$^{ \dagger}$ \qquad Athina P. Petropulu$^{\dagger}$ \qquad  H. Vincent Poor$^{\ddagger}$\thanks{Work supported by ARO  grant W911NF2110071 and NSF Grant CNS-2128448 }}

\address{$^{\dagger}$ Rutgers, The State University of New Jersey, Piscataway, NJ \\
$^{\ddagger}$Princeton University, Princeton, NJ}

\begin{document}
%\ninept
%

\maketitle
\begin{abstract}
\vspace{-0.3cm}
The paper studies the problem of designing the Intelligent Reflecting Surface (IRS) phase shifters for Multiple Input Single Output (MISO) communication systems in spatiotemporally correlated channel environments, where the destination can move within a confined area. The objective is to maximize the expected sum of SNRs at the receiver over infinite time horizons. The problem formulation gives rise to a Markov Decision Process (MDP). We propose a deep actor-critic algorithm that accounts for  channel correlations and  destination motion by constructing the state representation to include the current position of the receiver and the phase shift values and receiver positions that correspond to a window of previous time steps. The channel variability induces high frequency components on the spectrum of the underlying value function. We propose the preprocessing of the critic's input with a Fourier kernel which enables stable value learning. Finally, we investigate the use of the destination SNR as a component of the designed MDP state, which is common practice in previous work. We provide empirical evidence that, when the channels are spatiotemporally correlated, the inclusion of the SNR in the state representation interacts with function approximation in ways that inhibit convergence. 
\end{abstract}

\vspace{-0.4cm}
\section{Introduction}
\vspace{-0.4cm}
The emergence of Intelligent Reflecting Surfaces (IRSs) \cite{di2020smart} paves yet another avenue for impactful research in the area of Wireless Communications. IRS typically consists of a panel of passive reflective elements. Each element's reflection coefficient can be explicitly controlled independently of the coefficients of the other elements. Consequently, the deployment of IRS introduces a number of new degrees of freedom for wireless system design. Therefore, holds the promise for performance improvements in terms of Quality-of-Service (QoS) \cite{jiang2019over}, security \cite{yu2020robust}  and energy preservation \cite{ huang2019reconfigurable}.

For these ambitions to materialize, there needs to be a collective effort towards developing regimens that decide upon the state of the IRS phase shifters on the fly so as to imprint desirable conditions on the signal propagation environment. Typically, the phase shift design aspiration induces the formulation of NP hard problems. The authors in \cite{8855810, wu2019intelligent} propose IRS phase shift design approaches that are predicated upon semidefinite relaxations. The performance of said methods is compelling, but the methods induce high computational complexity ($\mathcal{O}(n^{6})$ where $n$ is the number of IRS phase shifters).

Due to the need for low computational complexity, researchers turned towards the employment of data-driven approaches for IRS phase-shift design. The authors in \cite{jiang2021learning} employ graph neural networks for estimating the IRS phase shifters. The neural network is trained, in a supervised fashion, to estimate the phase shift coefficients from channel pilots. The works in \cite{huang2019indoor, taha2021enabling} also pose the IRS phase shift design as a supervised learning problem and use labels to train a neural network to map channel measurements to phase shift values. Supervised learning methods require the availability of ground truth labels which are scarce for the problem at hand. Second, the supervised learning regimes assume that channel data are i.i.d samples from the same stationary distribution. This view is inadequate when the channels exhibit correlations with respect to time and space. This is mostly the case for urban communications due to shadowing \cite{wang2017stationarity}.

 Both shortcoming of supervised learning can be surpassed by the employment of deep Reinforcement Learning (RL). In the case of deep RL there is no need for labels, just measurements of a reward signal which can be the QoS metric of interest (SNR). Furthermore, the i.i.d assumption for the channel measurements can be lifted and the presupposition for spatiotemporal channel correlations can be readily adopted. Nonetheless such a setup has not been studied yet, to the best of our knowledge. The authors in \cite{feriani2021robustness} investigate a communication scenario between a MIMO source and a mobile destination, assisted by a single IRS. They benchmark the performance of a simple deep RL algorithm against multiple numerical methods under highly noisy channels. They demonstrate that the deep RL method is noticeably more robust with respect to noise. This provides strong incentive towards developing deep RL algorithms for IRS phase shift optimization. RL approaches have previously been employed for IRS phase shift design in single-user \cite{feng2020deep,feriani2021robustness} and multi-user \cite{huang2020reconfigurable,9766179} set ups. We differentiate from the aforementioned approaches in the following ways. First, all  previous methods do not assume the spatiotemporal correlations of the channels. We provide an algorithm that is specifically designed to address the case of spatiotemporally correlated channels. Channel correlations are very prominent in urban communication environments and arise due to shadowing.  Furthermore, the underlying value functions of MDPs that depend on spatiotemporally correlated channels typically possess high frequencies in the corresponding spectra \cite{9676432}. Recent results in deep learning regression demonstrate the inability of feedforward neural networks to represent high frequency functions \cite{cao2019towards} (a phenomenon coined as \textit{Spectral Bias}). Motivated by recent results in graphics \cite{tancik2020fourier}, we propose to preprocess the input of the critic with a Fourier kernel to overcome the Spectral Bias. Finally, we investigate the inclusion of the destination SNR in the representation of the MDP state. This has been common practice in previous work \cite{feng2020deep,feriani2021robustness, huang2020reconfigurable}. We showcase that, for spatiotemporally varying channels, including the SNR in the state interacts undesirably with function approximation and causes divergence and instability.
 \vspace{-0.2cm}
\begin{figure}[ht] 
\centering
{\label{fig:lossideal}\includegraphics[scale=0.13]{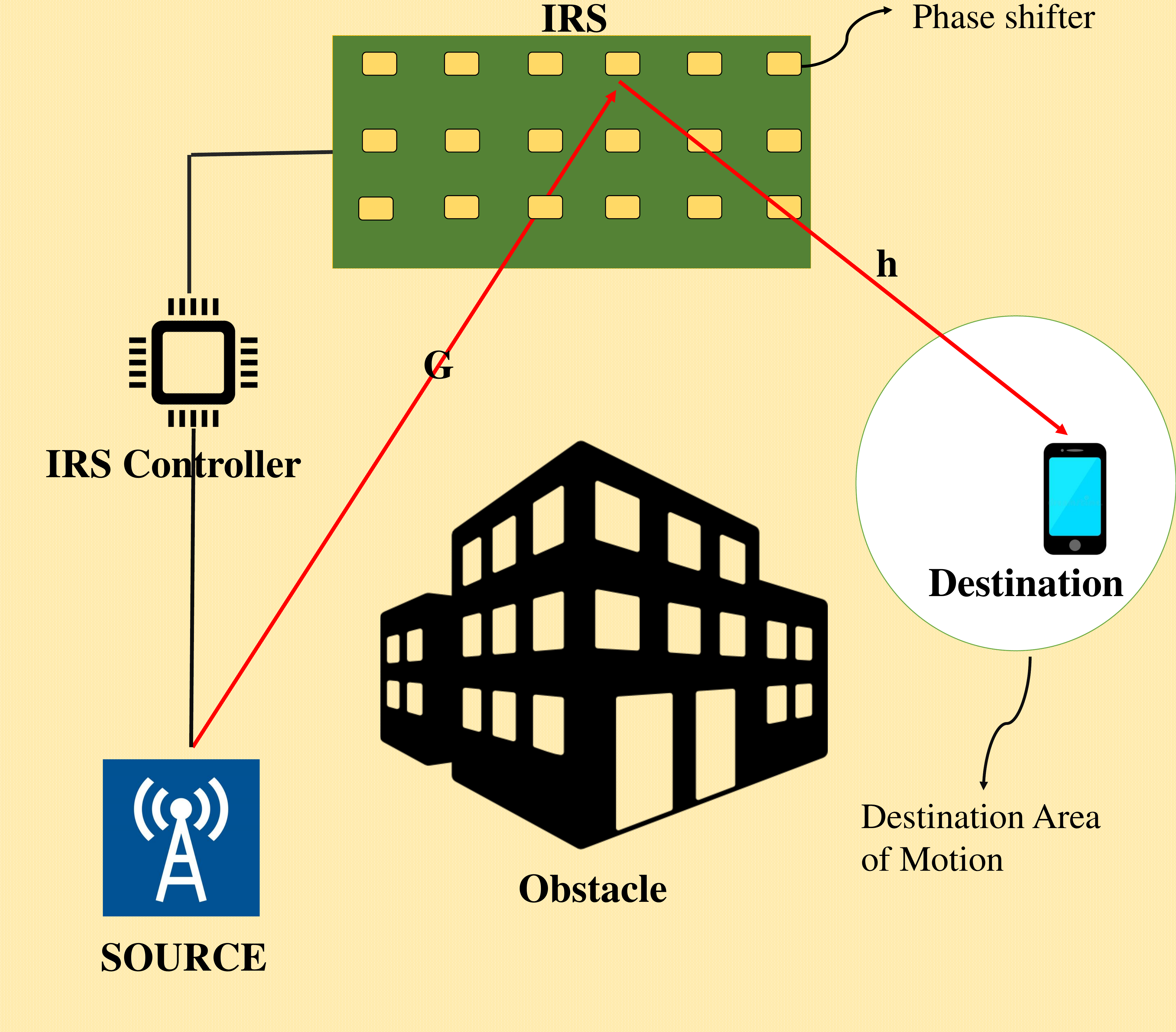}}%
\caption{\small IRS-aided single-user MISO} \label{MISO}
\vspace{-0.8cm}
\end{figure}
\section{Signal Model}
\vspace{-0.4cm}
We consider a Multiple-Input-Single-Output (MISO) Communication system, as shown in Fig. \ref{MISO}. The source is a Uniform Linear Array (ULA) with $N$ antennas. The IRS is a panel of $M = M_{x} \times M_{y}$ passive phase shift elements. The source-IRS and IRS-destination channels are denoted by  $\mathbf{G} \in \mathbb{C}^{M \times N}$, and  $\mathbf{h} \in \mathbb{C}^{M \times 1}$, respectively.
The IRS phase shifts are denoted by  $\theta_{i} \in [-\pi,\pi]$, $i=1, \dots M$. The source transmits symbol $\mu(t) \in \mathbb{C}$, precoded by $\mathbf{b} \in \mathbb{C}^{N \times 1}$, with  $\lVert \mathbf{b} \rVert^{2} \leq P_{max}$.
 We assume that there is no direct link from the source to the destination, so the  propagation path goes through the IRS. The received signal is
%  The source's beamforming weights are denoted as $\mathbf{b} \in \mathbb{C}^{N \times 1}$ and $\lVert \mathbf{b} \rVert^{2} \leq P_{max}$. 
\vspace{-0.2cm}
\begin{equation}
    y =  \mathbf{h}^{H} \mathbf{\Phi} \mathbf{G} \mathbf{b} \mu + n
    \vspace{-0.2cm}
\end{equation}
where $\mathbf{\Phi} = \text{diag}(e^{j \theta_{1}}, e^{j \theta_{2}}, \dots e^{j \theta_{M}})$, $n \sim \mathcal{C} \mathcal{N}\left(0, \sigma^{2}\right)$ is the reception noise at the destination and $\mathbb{E}\left[ \mu^{2} \right] = 1$.
If we assume that the phase shifters are fixed ($\mathbf{\Phi}$ is fixed), then the optimal beamforming weights are given by \cite{wu2019intelligent}:
\vspace{-0.2cm}
\begin{equation}
    \mathbf{b}^{*} = \sqrt{P_{max}} \frac{(\mathbf{h}^{H} \mathbf{\Phi} \mathbf{G}) ^ {H}}{\lVert \mathbf{h}^{H} \mathbf{\Phi} \mathbf{G} \rVert}
    \vspace{-0.3cm}
\end{equation}

We adopt the following assumptions. First of all, both channels (BS-IRS and IRS-destination) exhibit correlations with respect to time, so $\mathbf{h} = \mathbf{h}(t)$ and $\mathbf{G} = \mathbf{G}(t)$. We also assume that the destination can move, but its motion is confined to a small area of space. The channel from the IRS to the destination also exhibits correlations with respect to space that depend on the relative position of the destination to the IRS. So, $\mathbf{h} = \mathbf{h}(t,\mathbf{x}_t)$, where $\mathbf{x}_t \in \mathbb{R}^{3}$ is the position of the destination in the $3D$ space at time step $t$;
$\mathbf{x}_t$  is assumed known at the IRS controller, estimated by a coexistent
radar perception system \cite{li2022dual}.
One way to design the phase shift matrix $\mathbf{\Phi}(t)$ at time $t$, assuming that the channels $\mathbf{h}(t)$ and $\mathbf{G}(t, \mathbf{x}_{t}$ are known, is to solve the following problem:
\vspace{-0.3cm}
\begin{equation}
\begin{aligned}
\max_{\mathbf{\Phi}(t)} \quad &   SNR(t) \equiv \frac{1}{\sigma^{2}}{\left \lVert \mathbf{h}\left(t, \mathbf{x}_t\right)^{H} \mathbf{\Phi}(t) \mathbf{G}(t) \right \rVert^{2}} \\
 \textrm{s.t.} \quad & \left\lvert \mathbf{\Phi}(t)_{(i,i)} \right\rvert = 1, \hspace{0.2cm} \forall i = 1, 2, \dots , M.
\end{aligned} \label{problem formulation}
\vspace{-0.1cm}
\end{equation}
Even if one neglects the computational overhead of estimating the channels at each time step (which is not negligible in practice), solving problem \eqref{problem formulation} induces complexity $\mathcal{O}(M^{6})$ \cite{8855810}. The calculation cost of the solution at every time step is prohibitive for real system deployment.

Instead of solving the above problem, we design an actor-critic algorithm that learns a policy which maps the position of the destination at time step $t$ and \textit{information} from previous time steps to the IRS phase shift values of the current time step. The policy is constructed so as to maximize the expected sum of destination SNRs over infinite time horizons of system operation. In such case, after learning the policy/actor, the IRS phase shift computation is a simple forward pass through the policy function.
\vspace{-0.5cm}
\section{Off-Policy Deep RL}
\vspace{-0.4cm}
Reinforcement Learning (RL) studies the class of problems where an agent interacts with the environment in a sequence of states, actions and rewards. This gives birth to a Markov Decision Process (MDP). The MDP is defined as a tuple $ \langle S, A, R,P,p, \gamma \rangle $, where $S$ is the state space, $A$ is the action space, $R:S \times A \rightarrow \mathbb{R}$ is the reward function and $P:S \times A \rightarrow S$ is the transition function that depends on the environment dynamics. $p(\mathbf{s})$ is the initial state distribution and $\gamma \in (0,1)$ is the discount factor that quantifies how "far-sighted" the agent is. The goal of RL is to learn a mapping from states to actions, namely a policy, $\pi (\mathbf{a} | \mathbf{s})$ that maximizes the expected discounted sum of rewards $J = \mathbb{E} \Bigl[\sum_{t=0}^{\infty} \gamma^{t} R(\mathbf{s}_t, \mathbf{a}_t)\Bigr] = \mathbb{E} \Bigl[\sum_{t=0}^{\infty} \gamma^{t} r(t)\Bigr]$.

The state-action value function $Q^{\pi}$ is defined as the expected discounted sum of rewards starting from a state-action pair and following the policy $\pi$ thereafter. 
\vspace{-0.4cm}
\begin{equation}
Q^{\pi}(\mathbf{s},\mathbf{a}) = \mathbb{E} \Bigl[\sum_{t=0}^{\infty} \gamma^{t} R(\mathbf{s}_t, \mathbf{a}_t)| (\mathbf{s}_0, \mathbf{a}_0) = (\mathbf{s},\mathbf{a})\Bigr]
\vspace{-0.4cm}
\end{equation}
The optimal value function $Q^{*}(\mathbf{s},\mathbf{a})$ is the fixed point of the Bellman backup operator:
\vspace{-0.3cm}
\begin{equation}
B^{*}Q(\mathbf{s},\mathbf{a}) = \mathbb{E}_{\mathbf{s}' \sim P} \Bigl[R(\mathbf{s},\mathbf{a}) + \gamma \max_{\mathbf{a}'} Q(\mathbf{s}', \mathbf{a}')\Bigr]
\vspace{-0.2cm}
\end{equation}
In off-policy RL, we assume the availability of an Experience Replay that contains $N_{exp}$ transitions from the interaction of the agent with the environment $\left(D = \{\mathbf{s}_{i}, \mathbf{a}_{i}, \mathbf{s}_{i}', r_{i} \}_{i=1}^{N_{exp}}\right)$ .  The optimal state-action value function is parameterized as a neural network with parameters $\mathbf{w}$, denoted as $ Q_{\mathbf{w}}(\mathbf{s},\mathbf{a})$. The process of approximating the value function entails the sampling of a batch of transitions from the Experience Replay and the following parameter update:
\vspace{-0.3cm}
\begin{multline}
    \mathbf{w} \rightarrow \mathbf{w} + 2 \eta \mathbb{E}_{(\mathbf{s},\mathbf{a},\mathbf{s}',r) \sim D}\\\Bigl[ \Bigl (B^{*}Q_{\mathbf{w}}(\mathbf{s},\mathbf{a}) - Q_{\mathbf{w}}(\mathbf{s},\mathbf{a})\Bigr) \nabla_{\mathbf{w}}Q_{\mathbf{w}}(\mathbf{s},\mathbf{a})\Bigr] \label{update rule q}
    \vspace{-1.5cm}
\end{multline}
The process of learning the value function is usually unstable due to the fact that the bootstrapping target $\Bigl (B^{*}Q_{\mathbf{w}}(\mathbf{s},\mathbf{a}) \Bigr)$ depends on the estimator $Q_{\mathbf{w}}(\mathbf{s},\mathbf{a})$. One popular heuristic includes the use of a target network with parameters $\boldsymbol{\psi}$ for computing the bootstrapping target. The parameter vector $\boldsymbol{\psi}$ slowly tracks the parameter vector $\mathbf{w}$ as proposed in \cite{lillicrap2015continuous}.

The policy is  parameterized by a neural network with parameters $\boldsymbol{\phi}$, denoted as $\pi_{\boldsymbol{\phi}}(\mathbf{s})$. The policy is updated by the following  rule (deterministic policy gradient theorem):
\vspace{-0.2cm}
\begin{equation}
    \boldsymbol{\phi} \rightarrow \boldsymbol{\phi} +  \eta \mathbb{E}_{\mathbf{s} \sim D}\Bigl[ \nabla_{\mathbf{a}} Q_{\mathbf{w}}(\mathbf{s},\mathbf{a}) |_{\mathbf{a} = \pi_{\boldsymbol{\phi}}(\mathbf{s})} \nabla_{\boldsymbol{\phi}}\pi_{\boldsymbol{\phi}}(\mathbf{s}) \Bigr] \label{update rule pi}
    \vspace{-0.3cm}
\end{equation}

\section{Deep RL for IRS phase shift design}
\vspace{-0.3cm}
In order to develop the RL algorithm for the IRS phase shift design we need to define the elements of the MDP.

\underline{\textbf{State}}: The first component of the state is the position of the destination, $\mathbf{x}_t \in \mathbb{R}^{3}$. It can be estimated by a deployed radar perception system. Since the channels evolve as a temporally correlated process, the IRS phase shifters will also exhibit spatiotemporal correlations. \textbf{We, therefore, include the phase shift values of the IRS  for a \underline{window} of previous time steps along with the destination positions at these time steps}. The window length is denoted as $W$. The incentive is that, by including the phase shifters and destination locations for enough previous time steps, we can safely assume that the state is Markovian. In that sense, we operate in the context of a fully observable MDP. This nuance is similar to the practice in \cite{mnih2015human}, where the state is the concatenation of consecutive frames for the environments of the Atari Domain.
\vspace{-0.3cm}
    \begin{equation*}
        \mathbf{s}_{t} = \left [ \mathbf{x}_t, \mathbf{x}_{t-1}, \theta_{1}^{t-1}, \dots, \theta_{M}^{t-1}, \dots , \mathbf{x}_{t-W}, \dots ,\theta_{M}^{t-W} \right],
        \vspace{-0.3cm}
    \end{equation*}
where $\theta_{i}^{j}$ is the value of the IRS phase shifter $i$ at time step $j$.

\underline{\textbf{Action}}: The action is the component-wise difference between the phase shift elements of the current step and those of the previous step:
\vspace{-0.3cm}
\begin{equation*}
    \mathbf{a}_t = \left[ \delta \theta^{t}_{1}, \delta \theta^{t}_{2}, \dots \delta \theta^{t}_{M} \right],
    \vspace{-0.3cm}
\end{equation*}
where $\delta\theta^{t}_{i} = \theta_{i}^{t} - \theta_{i}^{t-1}$.

\underline{\textbf{Reward}}: The reward is the destination SNR at step $t$:
\vspace{-0.3cm}
\begin{equation*}
    r(t) = SNR(t).
    \vspace{-0.3cm}
\end{equation*}

We adopt the DDPG \cite{lillicrap2015continuous} algorithm to learn the policy that maps the agent's state to the phase shifters of the IRS. We employ $3$ main networks, namely the actor network, $\pi_{\phi}$, and the critic networks, $ Q_{w_{1}}$ and $ Q_{w_{2}}$. At every update step, we sample a batch of transitions from the Experience Replay and update the $2$ critic networks with the update rule of eq. \eqref{update rule q} and the actor with the update rule of eq. \eqref{update rule pi}. The $2$ critic networks are trained independently, but the minimum of the $2$ is used for the update of the policy (eq. \eqref{update rule pi}) to hedge against overestimation. We employ corresponding target networks for all the main ones and update their parameters following the framework in \cite{lillicrap2015continuous}.

Each neural network is a Multilayer Perceptron with ReLU activations (ReLU MLP). A critical question is how to ensure that the unit modulus constraints of problem \eqref{problem formulation} are satisfied during training. We address the challenge by explicitly designing the actor architecture. In particular,  the in-between-layer activations of the actor are ReLUs, but the activation of the output layer is hyperbolic tangent (Tanh). The Tanh squashes each output component to the $\left[-1, 1 \right]$ range. Subsequently, we multiply each output component with $\pi$. Therefore, we ensure that the output of the actor belongs to $\left[-\pi, \pi \right]^{M}$. Since the actor estimates the difference between the phase shifters of the current time step and those of the previous time step, we clip each resulting phase shifter to $\left[-\pi, \pi \right]$ for every IRS element at every time step. We denote this variation as \textbf{RL-IRS-Base}.

It has been common practice, when employing Deep RL for IRS phase shift design, to include the destination service metric (SNR) in the state representation \cite{feng2020deep,feriani2021robustness,huang2020reconfigurable}. Even though such a choice might seem intuitive, the way that it interacts with function approximation remains elusive. We implement a variation of the \textbf{RL-IRS-Base} that includes the destination SNR as a component of the state to test whether the aforementioned design choice is appropriate for the case of spatiotemporally correlated channels. We denote this variation as \textbf{RL-IRS-SNR-state}.
\vspace{-0.5cm}
\subsection{Fourier Preprocessing}
\vspace{-0.3cm}
%  Recent results in deep learning theory reveal a bias of ReLU MLPs towards low frequency spectra, a phenomenon called \textit{Spectral Bias} \cite{rahaman2019spectral}. 
 As discussed in \cite{9676432,9746407}, value functions that depend on spatiotemporally correlated communication channels typically inhere high frequency content. When it comes to the \textbf{RL-IRS-Base} algorithm, the critics are ReLU MLPs that regress the value function of the formulated MDP. The value function depends on spatiotemporally correlated channels. Due to spectral bias, the critic might not be able to learn the high frequencies of the value function. Since the policy is updated to select the action that maximizes the critic estimate at each state, inaccurate value estimation results in suboptimal policies and/or divergence. We propose the preprocessing of the state-action pair with a random Fourier Kernel before passing it through the critic. We denote as $\mathbf{\Tilde{s}} \in \mathbb{R}^{(W + 1)(M+3)}$ the concatenation of  state $\mathbf{s}$ and  action $\mathbf{a}$, i.e., 
\vspace{-0.2cm}
\begin{equation}
    \mathbf{\Tilde{s}}= [\mathbf{s},\mathbf{a}]^ \mathsf{T} \rightarrow \mathbf{v} = [cos(2\pi {\bf B} \mathbf{\mathbf{\Tilde{s}}}), sin(2\pi {\bf B} \Tilde{s})]^ \mathsf{T},
    \label{B}
    \vspace{-0.2cm}
\end{equation}
\noindent where  matrix ${\bf B}$ is the transformation Kernel. Each element of ${\bf B}$ is drawn from $\mathcal{N}(0,\sigma_{ {\bf B}}^2)$.  The operations $cos(\cdot)$ and $sin(\cdot)$ in \eqref{B} are applied element-wise. The variation that employs the Fourier preprocessing is denoted as \textbf{RL-IRS-FF}.
\vspace{-0.2cm}
\begin{figure}[ht] 
\centering
{\label{fig:lossideal}\includegraphics[scale=0.45]{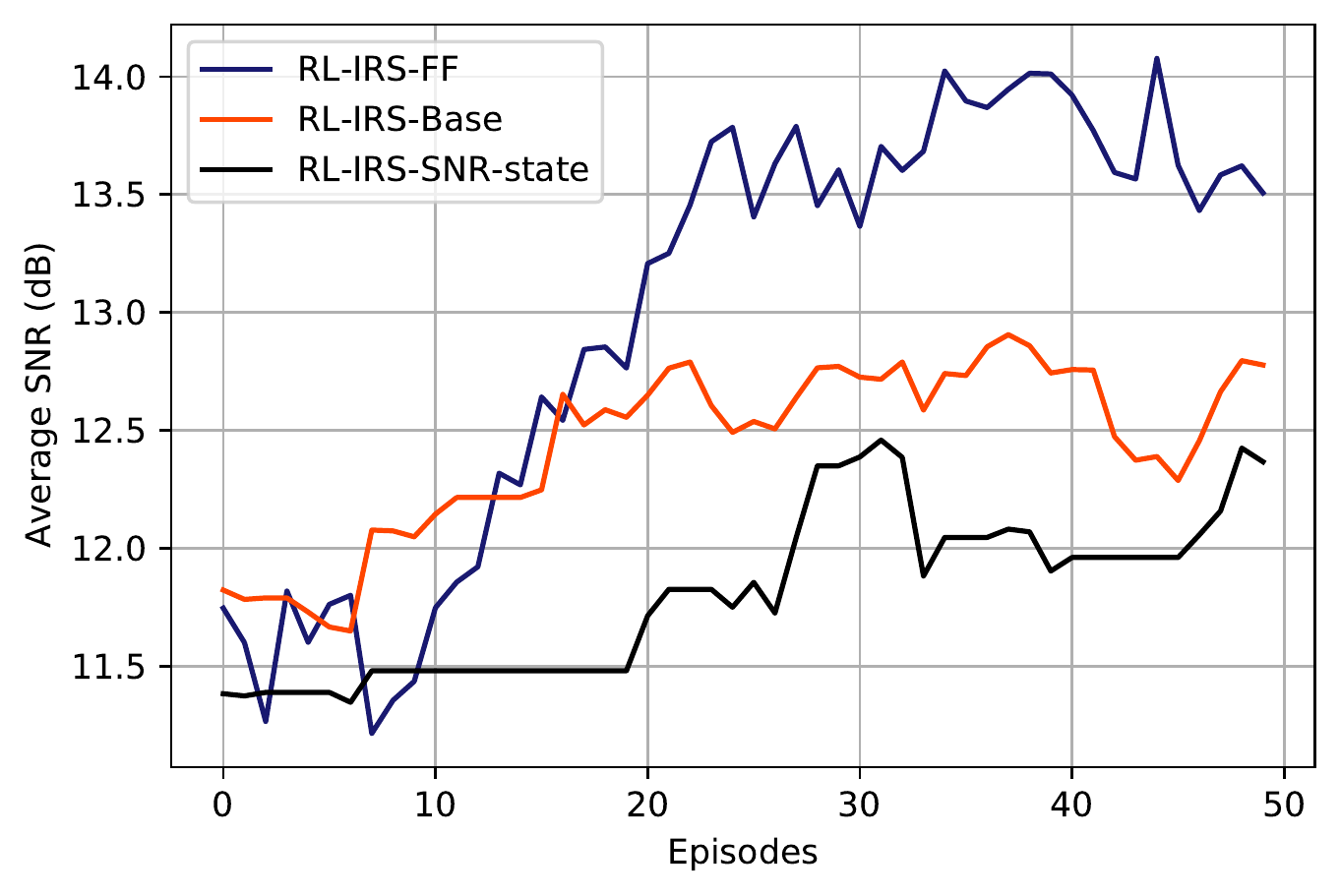}}%
\vspace{-0.4cm}
\caption{\small The training performance of the $3$ proposed algorithms for $50$ episodes. Each episode is comprised by $300$ time steps. Each curve is the average over $10$ seeds. }
\vspace{-0.7cm}
\label{comparison }
\end{figure}
\vspace{-0.5cm}
\section{Experiments}
\vspace{-0.4cm}
We simulate a scenario as in Fig. \ref{MISO}. The operation space is a $3$D cube $(20^{3} m^{3})$, discretized into cube cells of volume $(1 m^{3})$. The source and IRS positions are fixed. The destination can freely move in an area within $4$ cube cells. It occupies one cell per time slot and can also move to a neighboring cell at the subsequent slot. We simulate the  channels so that they have statistics as described in \cite{kalogerias2017spatially}. The log-magnitude of the channel has $3$ additive components, the pathloss with exponent $l=2.3$, the multipath, which is i.i.d zero-mean Gaussian with variance $\sigma_{\xi}=0.6$, and the shadowing which is a a zero-mean Gaussian correlated in time and space. The correlation distance is $c_1=1.2$, the correlation time $c_2=5$, and the shadowing power is $\eta^{2}=6$. The source transmit power is $P_{max}=65dBm$. The noise variance at the destination is $ \sigma^{2}=0.5$.  These parameters are consistent with real measurements \cite{wang2017stationarity} for urban channels. Each neural network is an MLP with $3$ layers. Each layer consists of $400$ neurons. We employ the Adam optimizer for the updates, with a learning rate of $2$e-$4$ and batch size of $64$. The elements of the Fourier kernel are drawn from a gaussian distribution with variance $0.01$. The Experience Replay size is $1$e+$6$ and the discount factor $\gamma$ is $0.99$. The source has $5$ antenna elements.
\vspace{-0.2cm}
\begin{figure}[ht] 
\centering
{\label{fig:lossideal}\includegraphics[scale=0.45]{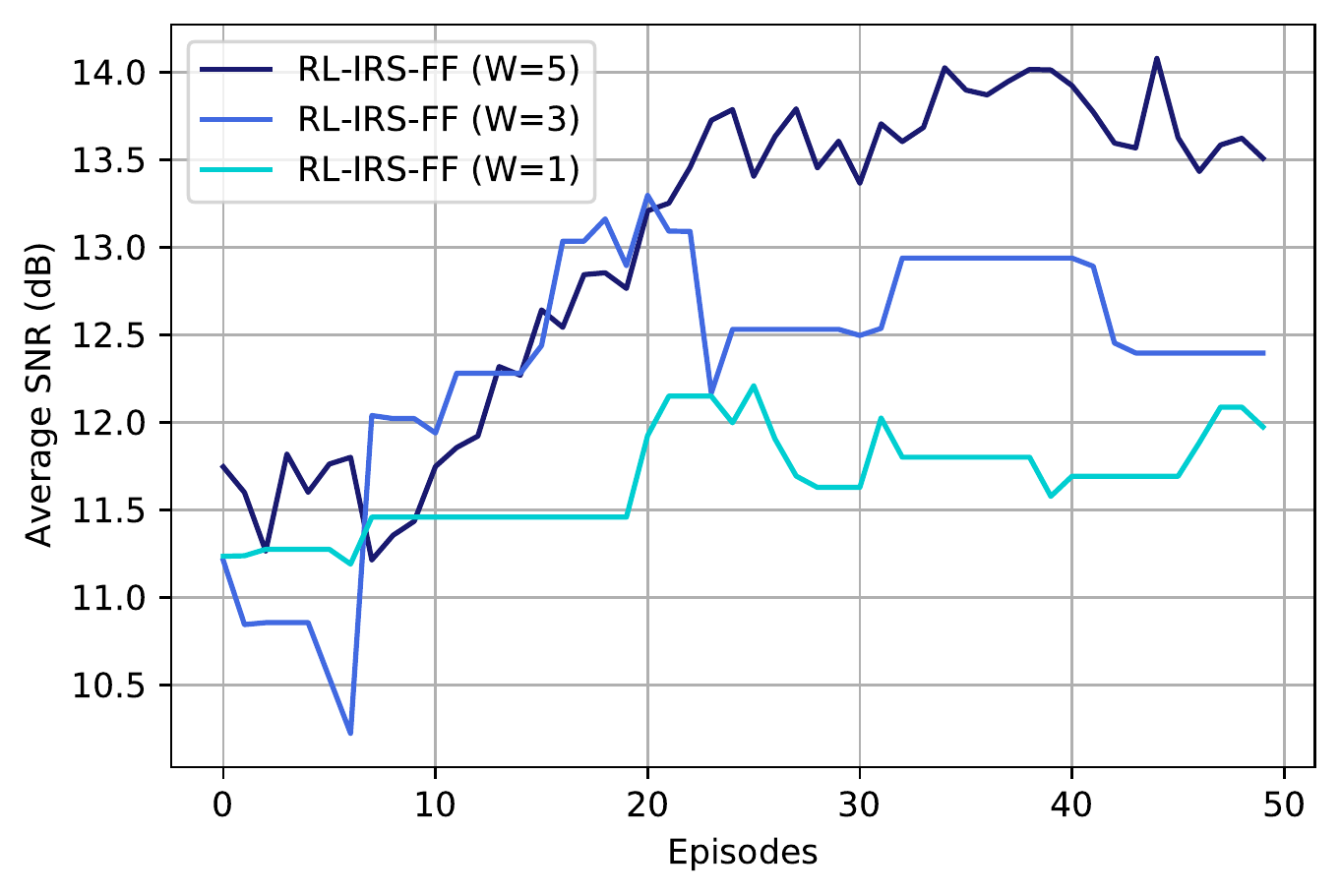}}%
\vspace{-0.4cm}
\caption{\small  The training performance of \textbf{RL-IRS-FF} for $3$ different values of the window size $W$. Each episode is comprised by $300$ time steps and each curve is an average over $10$ seeds.}
\vspace{-0.3cm}
\label{comparison w}
\end{figure}

\begin{figure}[ht] 
\centering
{\label{fig:lossideal}\includegraphics[scale=0.45]{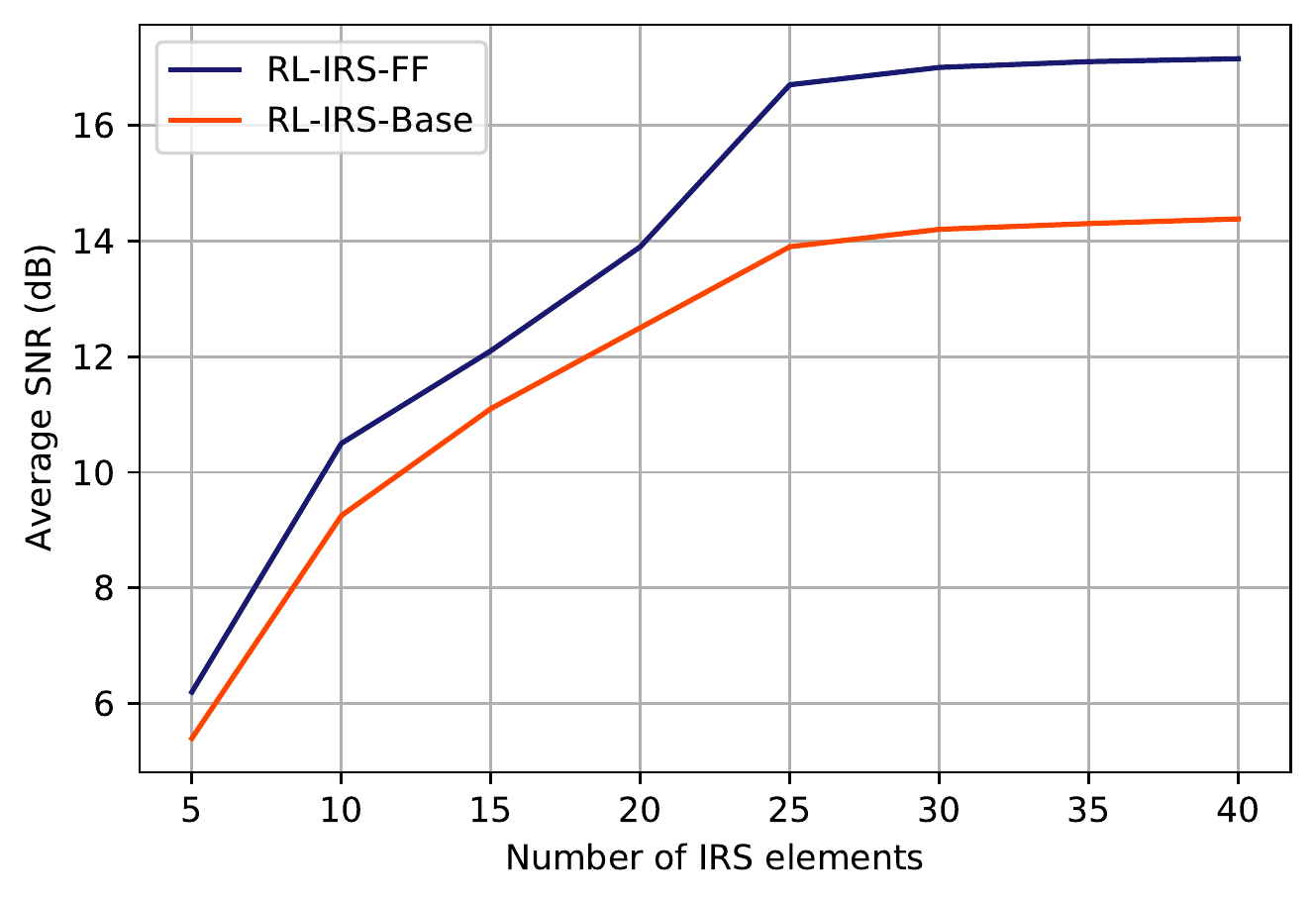}}%
\vspace{-0.4cm}
\caption{\small  The performances of \textbf{RL-IRS-FF} and \textbf{RL-IRS-Base}, after convergence, with respect to the size of the IRS.}
\vspace{-0.6cm}
\label{upon convergence}
\end{figure}
As can be inferred by Fig. \ref{comparison }, the algorithm that introduces the Fourier preprocessing performs significantly better than both the base implementation and the base implementation that includes the SNR at the state when it comes to convergence speed and destination SNR (increase of approximately $1.5dB$) . We attribute this to the amelioration of the Spectral Bias and, therefore, the more precise estimation of the value function. Fig. \ref{comparison w} demonstrates the performance of the \textbf{RL-IRS-FF} for $3$ different values of the window size $W$. The performance is improved for larger window sizes and best performance is exhibited for $W=5$ which is equal to the temporal correlation coefficient of the channels. We noticed that stability requires a window size that is, at least, equal to the coefficient of the temporal correlation. The experiments for Figures \ref{comparison } and \ref{comparison w} are conducted with an IRS of $20$ elements. Fig. \ref{upon convergence} illustrates the peformances of \textbf{RL-IRS-FF} and \textbf{RL-IRS-Base}, after convergence, with respect to the number of IRS elements. The Fourier features provide consistent improvements as we scale to larger IRSs. Both approaches seem to plateau at about $25$ elements. At this point, we have probably reached the representation capacity of the critic. Further increase in performance for IRSs of larger sizes would require an increase in the parameter size of the networks (critics and actor). The variation that includes the SNR as a state component is prone to divergence. We attribute this to the fact that, when the channels are correlated, states that correspond to different destination positions and different IRS phase shift values for a window $W$ of previous time steps can result in similar SNRs. Therefore, by including the SNR in the state, we implicitly cause \textit{state-aliasing}, where intrinsically different states are processed by the critic as similar. This agrees with results in neural value approximation \cite{achiam2019towards} which report that state-aliasing causes instability during value learning.
\vspace{-0.4cm}
\section{Conclusion}
\vspace{-0.4cm}
The paper proposes a deep actor-critic algorithm for IRS phase shift design for MISO communication systems and spatiotemporally correlated channels. The variability of the correlated channels induces high frequency components on the spectrum of the underlying value function. We  propose the preprocessing of the state-action vector with a Fourier Kernel before passing it through the critic, which helps in the process of learning the high frequencies of the value function. We empirically showcase that the inclusion of the SNR in the state, a common practice in prior works, inhibits stability and convergence.  Further investigation on how the SNR as a state component interacts with value function approximation will be part of future work.  

\printbibliography

@article{jiang2021learning,
  title={Learning to reflect and to beamform for intelligent reflecting surface with implicit channel estimation},
  author={Jiang, Tao and Cheng, Hei Victor and Yu, Wei},
  journal={IEEE Journal on Selected Areas in Communications},
  volume={39},
  number={7},
  pages={1931--1945},
  year={2021},
  publisher={IEEE}
}

@ARTICLE{9676432,
  author={Evmorfos, Spilios and Diamantaras, Konstantinos I. and Petropulu, Athina P.},
  journal={IEEE Transactions on Signal Processing}, 
  title={Reinforcement Learning for Motion Policies in Mobile Relaying Networks}, 
  year={2022},
  volume={70},
  number={},
  pages={850-861},
  doi={10.1109/TSP.2022.3141305}}

@INPROCEEDINGS{9746407,
  author={Evmorfos, Spilios and Petropulu, Athina P.},
  booktitle={ICASSP 2022 - 2022 IEEE International Conference on Acoustics, Speech and Signal Processing (ICASSP)}, 
  title={Deep Actor-Critic for Continuous 3D Motion Control in Mobile Relay Beamforming Networks}, 
  year={2022},
  volume={},
  number={},
  pages={5353-5357},
  doi={10.1109/ICASSP43922.2022.9746407}}

@article{wu2019intelligent,
  title={Intelligent reflecting surface enhanced wireless network via joint active and passive beamforming},
  author={Wu, Qingqing and Zhang, Rui},
  journal={IEEE Transactions on Wireless Communications},
  volume={18},
  number={11},
  pages={5394--5409},
  year={2019},
  publisher={IEEE}
}

@inproceedings{huang2019indoor,
  title={Indoor signal focusing with deep learning designed reconfigurable intelligent surfaces},
  author={Huang, Chongwen and Alexandropoulos, George C and Yuen, Chau and Debbah, M{\'e}rouane},
  booktitle={2019 IEEE 20th international workshop on signal processing advances in wireless communications (SPAWC)},
  pages={1--5},
  year={2019},
  organization={IEEE}
}

@article{taha2021enabling,
  title={Enabling large intelligent surfaces with compressive sensing and deep learning},
  author={Taha, Abdelrahman and Alrabeiah, Muhammad and Alkhateeb, Ahmed},
  journal={IEEE access},
  volume={9},
  pages={44304--44321},
  year={2021},
  publisher={IEEE}
}

@inproceedings{wang2017stationarity,
  title={Stationarity region of mm-wave channel based on outdoor microcellular measurements at 28 GHz},
  author={Wang, Rui and Bas, Celalettin Umit and Sangodoyin, Seun and Hur, Sooyoung and Park, Jeongho and Zhang, Jianzhong and Molisch, Andreas F},
  booktitle={MILCOM 2017-2017 IEEE Military Communications Conference (MILCOM)},
  pages={782--787},
  year={2017},
  organization={IEEE}
}

@article{feng2020deep,
  title={Deep reinforcement learning based intelligent reflecting surface optimization for MISO communication systems},
  author={Feng, Keming and Wang, Qisheng and Li, Xiao and Wen, Chao-Kai},
  journal={IEEE Wireless Communications Letters},
  volume={9},
  number={5},
  pages={745--749},
  year={2020},
  publisher={IEEE}
}

@article{huang2020reconfigurable,
  title={Reconfigurable intelligent surface assisted multiuser MISO systems exploiting deep reinforcement learning},
  author={Huang, Chongwen and Mo, Ronghong and Yuen, Chau},
  journal={IEEE Journal on Selected Areas in Communications},
  volume={38},
  number={8},
  pages={1839--1850},
  year={2020},
  publisher={IEEE}
}

@article{tancik2020fourier,
  title={Fourier features let networks learn high frequency functions in low dimensional domains},
  author={Tancik, Matthew and Srinivasan, Pratul and Mildenhall, Ben and Fridovich-Keil, Sara and Raghavan, Nithin and Singhal, Utkarsh and Ramamoorthi, Ravi and Barron, Jonathan and Ng, Ren},
  journal={Advances in Neural Information Processing Systems},
  volume={33},
  pages={7537--7547},
  year={2020}
}

@article{di2020smart,
  title={Smart radio environments empowered by reconfigurable intelligent surfaces: How it works, state of research, and the road ahead},
  author={Di Renzo, Marco and Zappone, Alessio and Debbah, Merouane and Alouini, Mohamed-Slim and Yuen, Chau and De Rosny, Julien and Tretyakov, Sergei},
  journal={IEEE journal on selected areas in communications},
  volume={38},
  number={11},
  pages={2450--2525},
  year={2020},
  publisher={IEEE}
}

@article{huang2019reconfigurable,
  title={Reconfigurable intelligent surfaces for energy efficiency in wireless communication},
  author={Huang, Chongwen and Zappone, Alessio and Alexandropoulos, George C and Debbah, M{\'e}rouane and Yuen, Chau},
  journal={IEEE Transactions on Wireless Communications},
  volume={18},
  number={8},
  pages={4157--4170},
  year={2019},
  publisher={IEEE}
}

@inproceedings{jiang2019over,
  title={Over-the-air computation via intelligent reflecting surfaces},
  author={Jiang, Tao and Shi, Yuanming},
  booktitle={2019 IEEE Global Communications Conference (GLOBECOM)},
  pages={1--6},
  year={2019},
  organization={IEEE}
}

@article{yu2020robust,
  title={Robust and secure wireless communications via intelligent reflecting surfaces},
  author={Yu, Xianghao and Xu, Dongfang and Sun, Ying and Ng, Derrick Wing Kwan and Schober, Robert},
  journal={IEEE Journal on Selected Areas in Communications},
  volume={38},
  number={11},
  pages={2637--2652},
  year={2020},
  publisher={IEEE}
}

@article{li2022dual,
  title={Dual-Function Radar-Communication System Aided by Intelligent Reflecting Surfaces},
  author={Li, Yikai and Petropulu, Athina},
  journal={arXiv preprint arXiv:2204.04721},
  year={2022}
}

@article{lillicrap2015continuous,
  title={Continuous control with deep reinforcement learning},
  author={Lillicrap, Timothy P and Hunt, Jonathan J and Pritzel, Alexander and Heess, Nicolas and Erez, Tom and Tassa, Yuval and Silver, David and Wierstra, Daan},
  journal={arXiv preprint arXiv:1509.02971},
  year={2015}
}

@article{kalogerias2017spatially,
  title={Spatially Controlled Relay Beamforming: $2 $-Stage Optimal Policies},
  author={Kalogerias, Dionysios S and Petropulu, Athina P},
  journal={arXiv preprint arXiv:1705.07463},
  year={2017}
}

@article{cao2019towards,
  title={Towards understanding the spectral bias of deep learning},
  author={Cao, Yuan and Fang, Zhiying and Wu, Yue and Zhou, Ding-Xuan and Gu, Quanquan},
  journal={arXiv preprint arXiv:1912.01198},
  year={2019}
}

@article{feriani2021robustness,
  title={On the robustness of deep reinforcement learning in IRS-aided wireless communications systems},
  author={Feriani, Amal and Mezghani, Amine and Hossain, Ekram},
  journal={arXiv preprint arXiv:2107.08293},
  year={2021}
}

@article{achiam2019towards,
  title={Towards characterizing divergence in deep q-learning},
  author={Achiam, Joshua and Knight, Ethan and Abbeel, Pieter},
  journal={arXiv preprint arXiv:1903.08894},
  year={2019}
}

@INPROCEEDINGS{8855810,
  author={Yu, Xianghao and Xu, Dongfang and Schober, Robert},
  booktitle={2019 IEEE/CIC International Conference on Communications in China (ICCC)}, 
  title={MISO Wireless Communication Systems via Intelligent Reflecting Surfaces : (Invited Paper)}, 
  year={2019},
  volume={},
  number={},
  pages={735-740},
  doi={10.1109/ICCChina.2019.8855810}}

@article{mnih2015human,
  title={Human-level control through deep reinforcement learning},
  author={Mnih, Volodymyr and Kavukcuoglu, Koray and Silver, David and Rusu, Andrei A and Veness, Joel and Bellemare, Marc G and Graves, Alex and Riedmiller, Martin and Fidjeland, Andreas K and Ostrovski, Georg and others},
  journal={nature},
  volume={518},
  number={7540},
  pages={529--533},
  year={2015},
  publisher={Nature Publishing Group}
}

@ARTICLE{9766179,
  author={Zhong, Canwei and Cui, Miao and Zhang, Guangchi and Wu, Qingqing and Guan, Xinrong and Chu, Xiaoli and Poor, H. Vincent},
  journal={IEEE Transactions on Communications}, 
  title={Deep Reinforcement Learning-Based Optimization for IRS-Assisted Cognitive Radio Systems}, 
  year={2022},
  volume={70},
  number={6},
  pages={3849-3864},
  doi={10.1109/TCOMM.2022.3171837}}
\end{document}